\documentclass[pre, preprint,groupedaddress, showkeys,showpacs]{revtex4}

\usepackage{graphicx}
\usepackage{fancyhdr}
\begin{document}


\title{Transport in periodic potentials induced by fractional Gaussian noise}


\author{Bao-quan  Ai$^{1}$}
\author{Ya-feng He$^{2}$}
 \author{Wei-rong Zhong$^{3}$}

\affiliation{$^{1}$Laboratory of Quantum Information Technology,
ICMP and
 SPTE, South China Normal University, 510006 Guangzhou, China.\\
 $^{2}$College of Physics Science and
Technology, Hebei University, 071002 Baoding, China.\\
$^{3}$Department of Physics, College of Science and Engineering,
Jinan University, 510632 Guangzhou, China.}


\date{\today}
\begin{abstract}
\indent  Directed transport of overdamped Brownian particles driven
by fractional Gaussian noises is investigated in asymmetrically
periodic potentials. By using Langevin dynamics simulations, we find
that rectified currents occur in the absence of any external driving
forces. Unlike white Gaussian noises, fractional Gaussian noises can
break thermodynamical equilibrium and induce directed transport.
Remarkably, the average velocity for persistent fractional noise is
opposite to that for anti-persistent fractional noise. The velocity
increases monotonically with Hurst exponent for the persistent case,
whereas there exists an optimal value of Hurst exponent at which the
velocity takes its maximal value for the anti-persistent case.
\end{abstract}

\pacs{05. 40. Fb, 02. 50. Ey, 05. 40. -a}



\maketitle
\section {Introduction}
\indent In systems possessing spatial or dynamical symmetry
breaking, Brownian motion combined with unbiased external input
signals, deterministic or random alike, can assist directed motion
of particles in nanoscale systems\cite{a1}. The subject of the
fluctuation-induced transport was motivated by the challenge to
explain unidirectional transport in biological systems\cite{a2}, as
well as their potential technological applications ranging from
classical non-equilibrium models\cite{a3} to quantum
systems\cite{a4}. Ratchets have been proposed to model the
unidirectional motion driven by zero-mean non-equilibrium
fluctuations. Broadly speaking, ratchet devices fall into three
categories depending on how the applied perturbation couples to the
substrate asymmetry:  rocking ratchets\cite{a5}, flashing
ratchets\cite{a6}, and correlation ratchets \cite{a7}. Additionally,
entropic ratchets, in which Brownian particles move in a confined
structure, instead of a potential, were also extensively
studied \cite{a8}.\\
\indent Anomalous diffusion has attracted growing attention, being
observed in various fields of physics and related sciences
\cite{a9}, where by contrast with Brownian motion, long-range
temporal correlations induce nonstandard dynamical behaviors. The
diffusion is characterized through the power law form of the
mean-square displacement $\langle x^{2}(t)\rangle\propto
t^{\alpha}$. According to the value of the index $\alpha$, one can
distinguish subdiffusion ($0<\alpha<1$), normal diffusion
($\alpha=1$) and superdiffusion ($\alpha>1$). In the literature, two
popular stochastic models have been used to account for anomalous
diffusion. The first model is the continuous-time random walk
\cite{a9, a10, a11,a12, a13, a14}. In this model, the subdiffusion
is caused by the long waiting time between successive jumps and the
superdiffusion is induced by the long jumps. In the minimal L\'{e}vy
ratchet\cite{a10,a11,a12,a13,a14}, the heavy-tailed distribution of
the $\alpha$-stable noise can break the thermodynamical equilibrium and induce directed transport.\\
\indent The second model is fractional Brownian motion (FBM)
introduced by Mandelbrot and Van Ness \cite{a15}. FBM has wide
applications in some complex systems, such as monomer diffusion in a
polymer chain \cite{a16}, diffusion of biopolymers in the crowded
environment \cite{a17}, single file diffusion \cite{a18}, and
translocation of the polymer passing through a pore \cite{a19}. The
statistical properties of FBM are characterized by the Hurst
exponent $0<H<1$. In particular, its meansquared displacement
satisfies $\langle x^{2}(t)\rangle\propto t^{2H}$, thus for $H<1/2$
one can obtain the subdiffusive dynamics, whereas for $H>1/2$ the
superdiffusive one \cite{a20}.  In the last few years, there has
been a growing interest in the study of the FBM
\cite{a21,a22,a23,a24}. However, most studies of FBM focus on the
free FBM and a few studies on FBM have been involved the potentials.
Recently, Sliusarenko and coworkers \cite{a25} studied the escape
from a potential well driven by fractional Gaussian noises and found
that the escape becomes faster for decreasing values of Hurst
exponent. Chaudhury and Cherayila \cite{a26} studied the first
passage time distribution for barrier crossing in a double well
under fractional Gaussian noises. It is uncertain whether fractional
Gaussian noise can induce directed transport in the absence of any
external driving forces. In order to answer this question, we
studied the transport of overdamped Brownian particles driven by
fractional Gaussian noises in asymmetrically periodic potentials. We
focus on finding the rectified mechanism and how noise intensity and
Hurst exponent affect the transport.

\section{Model and Methods}
\indent In this study, we consider the directed transport of the
Brownian particles driven by fractional Gaussian noises in the
absence of whatever additional time-dependent forces. The overdamped
dynamics can be described by the following Langevin equation in the
dimensionless form
\begin{equation}\label{}
    \eta\frac{dx}{dt}=-U^{'}(x)+\sqrt{\eta k_{B} T}\xi_{H}(t),
\end{equation}
where $\xi_{H}(t)$ is the fractional Gaussian noise,
$D=\frac{k_{B}T}{\eta}$ is noise intensity,  and $H$ is the Hurst
exponent. $\eta$ is the friction coefficient of the particle,
$k_{B}$ is the Boltzmann constant, and $T$ is the absolute
temperature. The prime stands for differentiation over $x$. $U(x)$
is an asymmetrically periodic potential
\begin{equation}\label{}
    U(x)=-U_{0}[\sin(x)+\frac{\Delta}{4}\sin(2x)],
\end{equation}
where $U_{0}$ denotes the height of the potential and $\Delta$ is
its asymmetric parameter.

   \indent Fractional Gaussian noise is a zero mean stationary random
   process with long memory effects \cite{a21,a22,a23,a24,a25}. It is closely related to the
   FBM process \cite{b1}, which is defined
   as a Gaussian process with an exponent $0<H<1$ and
   \begin{equation}\label{}
   \langle\xi_{H}(t)\rangle=0,
   \end{equation}
   \begin{equation}\label{}
    \langle\xi_{H}(t)\xi_{H}(s)\rangle=\frac{1}{2}[t^{2H}+s^{2H}-(t-s)^{2H}],
   \end{equation}
for $0< s\leq t$. In the long time limit, the autocorrelation
   function will decay as
   \begin{equation}\label{}
    \langle\xi_{H}(0)\xi_{H}(t)\rangle \propto 2H(2H-1)t^{2H-2},
   \end{equation}
for $0<H<1$ and $H\neq \frac{1}{2}$. When $H=\frac{1}{2}$,
fractional Gaussian noise reduces to white Gaussian noise. From  Eq.
(5), it is easy to find that the  noises are positively correlated
(persistent case) for $\frac{1}{2}<H<1$, and negatively correlated
(anti-persistent case) for $0<H<\frac{1}{2}$.

 \indent Though FBM is an old topic, the
 consistent analytical methods are still not available. Here we will
 study the transport of the Brownian particles by using Langevin
 dynamics simulations. From Eqs. (1) and (4)
one can obtain the discrete time representation of Eq. (1) for
sufficiently small time step $\delta t$
\begin{equation}\label{}
    x(t_{n+1})=x(t_{n})-U^{'}(x(t_{n}))\delta t+\sqrt{D}\delta
    t^{H}\xi_{H}(n),
\end{equation}
where $n=0,1,2...$ and $\xi_{H}(n)$ is fractional Gaussian random
number.  We used the method described in\cite{a25,b2,b3} for
simulating fractional Gaussian random number.

\indent In this study, we mainly focus on the transport of the
driven particles. The average velocity $\upsilon$ is used to measure
the transport,
            \begin{equation}\label{}
            \upsilon(t)=\frac{1}{N}\sum_{i=1}^{N}\frac{x_{i}(t)-x_{i}(t_{0})}{t-t_{0}},
            \end{equation}
where $N$ is the number of the realizations and $t_{0}$ and $t$ are
the initial and the end time for the simulations, respectively. The
asymptotic velocity $V$ is
\begin{equation}\label{}
V=\lim_{t\rightarrow\infty}\upsilon(t).
\end{equation}

\section {Numerical results and discussion}
\begin{figure}[htbp]
  \begin{center}\includegraphics[width=8cm,height=7cm]{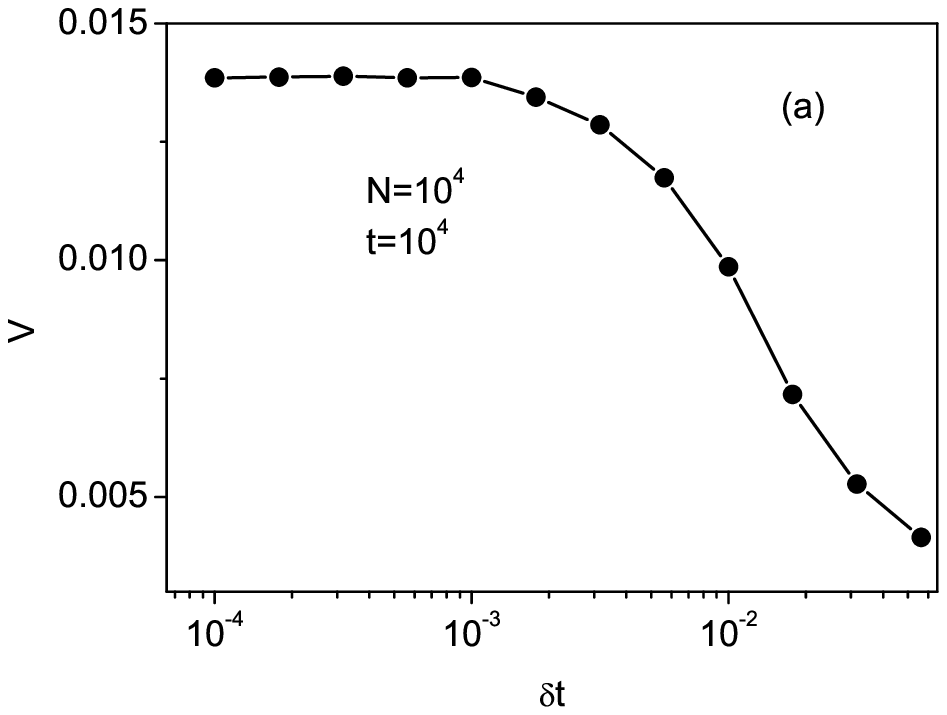}
  \includegraphics[width=8cm,height=7cm]{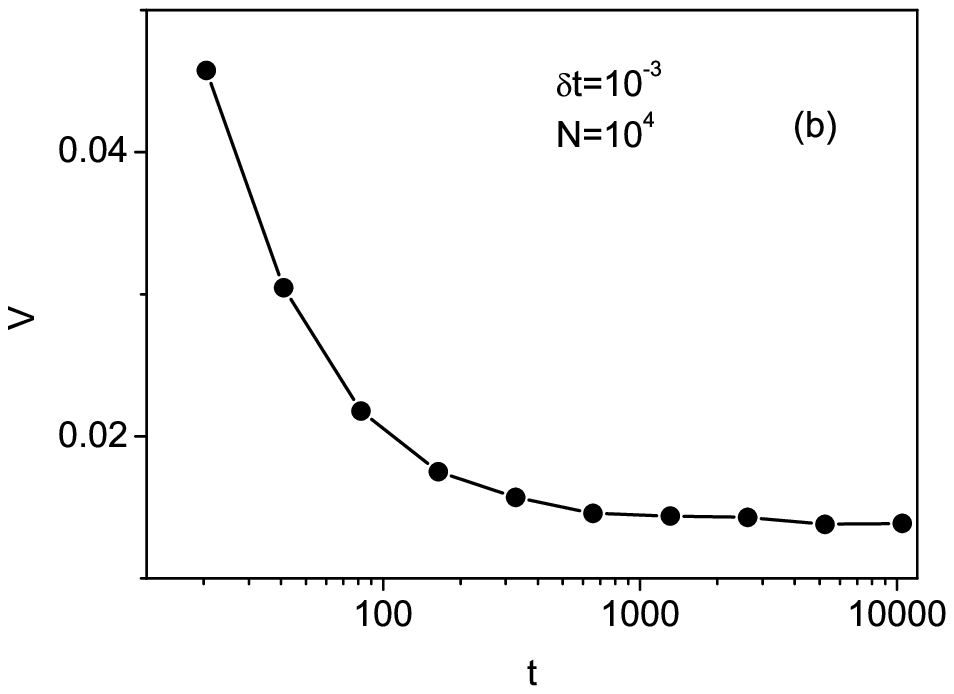}
  \includegraphics[width=8cm,height=7cm]{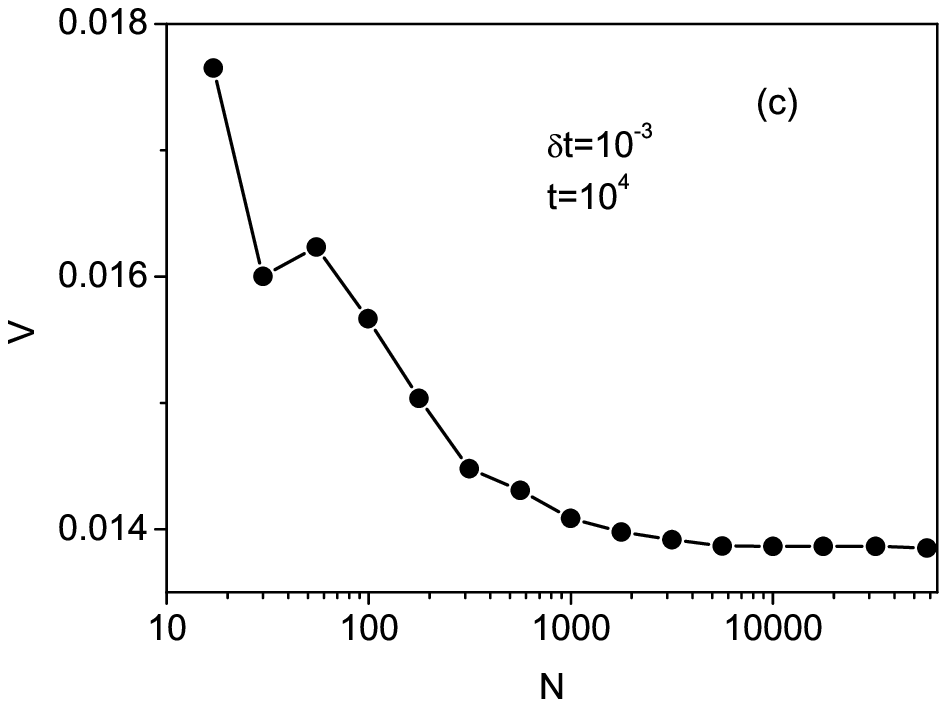}
  \includegraphics[width=8cm,height=7cm]{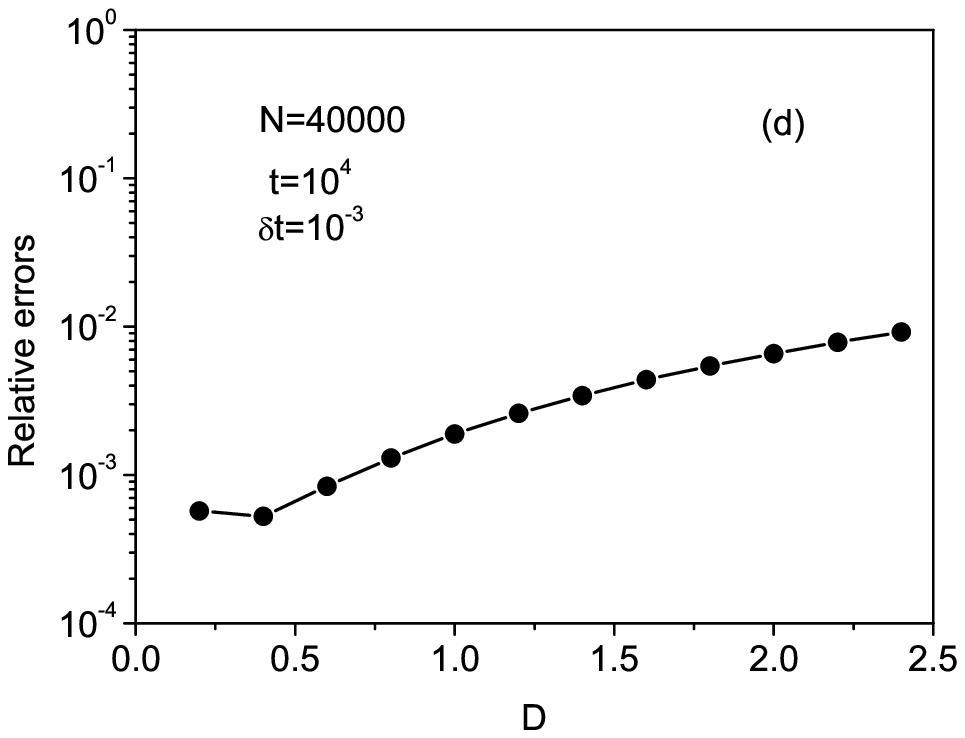}
  \caption{Convergence and relative errors of the algorithm. (a) Dependence of $V$ on time step $\delta t$ at $N=10^{4}$ and $t=10^{4}$;
  (b)dependence of $V$ on the end time $t$ for simulations
  at $\delta t=10^{-3}$ and $N=10^{4}$; (c) dependence of $V$ on the number $N$ of the
realizations at $\delta t=10^{-3}$ and $t=10^{4}$; (d) dependence of
estimated relative errors on noise intensity $D$.  The other
parameters are $D=0.5$, $U_{0}=1.0$, and $\Delta=1.0$.}\label{1}
\end{center}
\end{figure}

\indent In order to check the convergence of the algorithm, we have
studied the dependence of average  velocity on time step $\delta t$,
the end time $t$, and the number $N$ of the realizations. From Fig.
1(a), (b) and (c), we can see that the algorithm is convergent and
the numerical results do not depend on the calculation parameters
($\delta t$, $t$, and $N$) when $\delta t<10^{-3}$, $t>10^{4}$, and
$N>10^{4}$. Therefore, in our simulations, the number of the
realizations is more than $4\times10^{4}$ realizations, time step is
chosen to be smaller than $10^{-3}$ and $t=10^{4}$. Fig. 1(d) shows
the estimated relative errors as a function of the noise intensity
$D$ at $N=4\times 10^{4}$, $t=10^{4}$, and $\delta t=10^{-3}$. It is
found that the relative errors are less than $0.01$ even for large
values of the noise intensity. Therefore, the parameter we used are
sufficient to obtain consistent results.

\begin{figure}[htbp]
  \begin{center}\includegraphics[width=10cm,height=8cm]{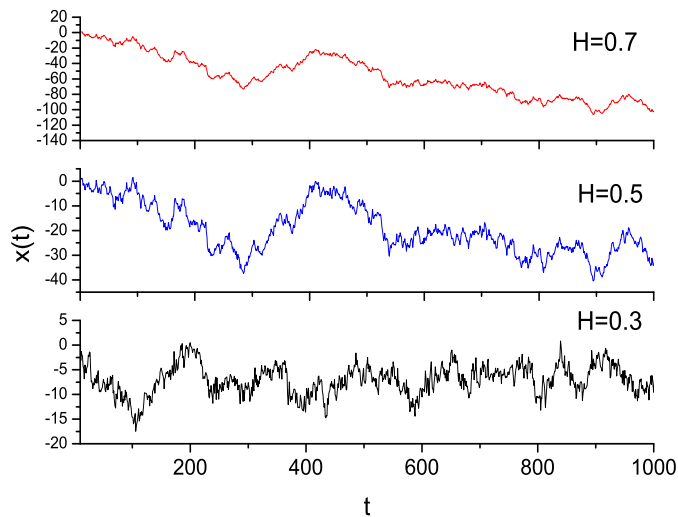}
  \caption{(Color online) Samples of FBM for different values of the Hurst exponent $H=0.3$, $0.5$, and $0.7$.}\label{1}
\end{center}
\end{figure}
\indent First, we study the  properties of FBM for both persistent
and anti-persistent cases. Figure 2 shows the simulated sample paths
for different values of $H$. The differences among these three cases
are clear. For $H=0.3$, the negative correlation accounts for high
variability, whereas the sample is more smooth for $H=0.7$ due to
the positive correlation \cite{b3}.

\begin{figure}[htbp]
  \begin{center}\includegraphics[width=10cm,height=8cm]{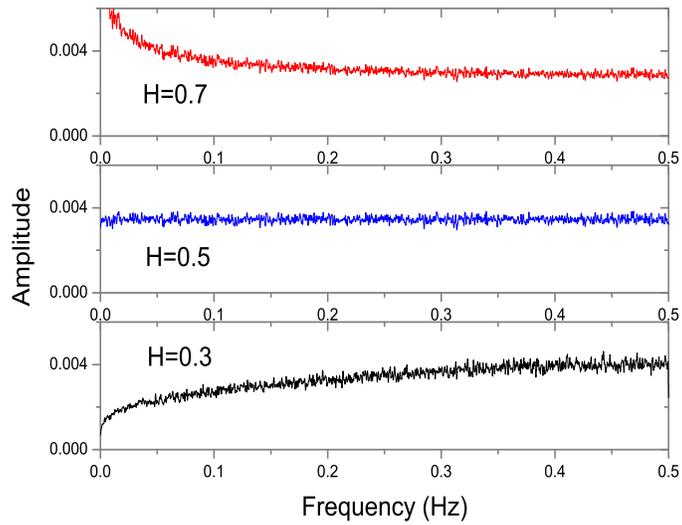}
  \caption{(Color online) The spectral density of the noises for different values of Hurst exponent $H=0.3$, $0.5$, and $0.7$.}\label{1}
\end{center}
\end{figure}
\indent The frequency spectrum of the external drive is very
important to determine direction of motion of the Brownian particles
in periodic potentials. In our minimal ratchet setup, fractional
Gaussian noise is the only external drive, so it is necessary to
analyze its frequency properties. In Fig. 3, we investigate the
spectral density of fractional Gaussian noise for different values
of $H$. We can find that the distributions of the frequency are
different for the three cases. For white Gaussian noise ($H=0.5$),
the spectral density is uniform. However, the low frequency
component is larger than the high frequency part in the spectral
density for the persistent case ($H=0.7$).  For the anti-persistent
case ($H=0.3$), the high frequency component is larger in the
spectral density. In order to facilitate the analysis of the driving
mechanisms, persistent fractional Gaussian noise can be artificially
divided into two frequency components: white Gaussian noise and low
frequency ac drive. Similarly, the anti-persistent fractional
Gaussian noise in frequency domain is equivalent to a compound of
white Gaussian noise and high frequency drive.

\begin{figure}[htbp]
  \begin{center}\includegraphics[width=8cm,height=8cm]{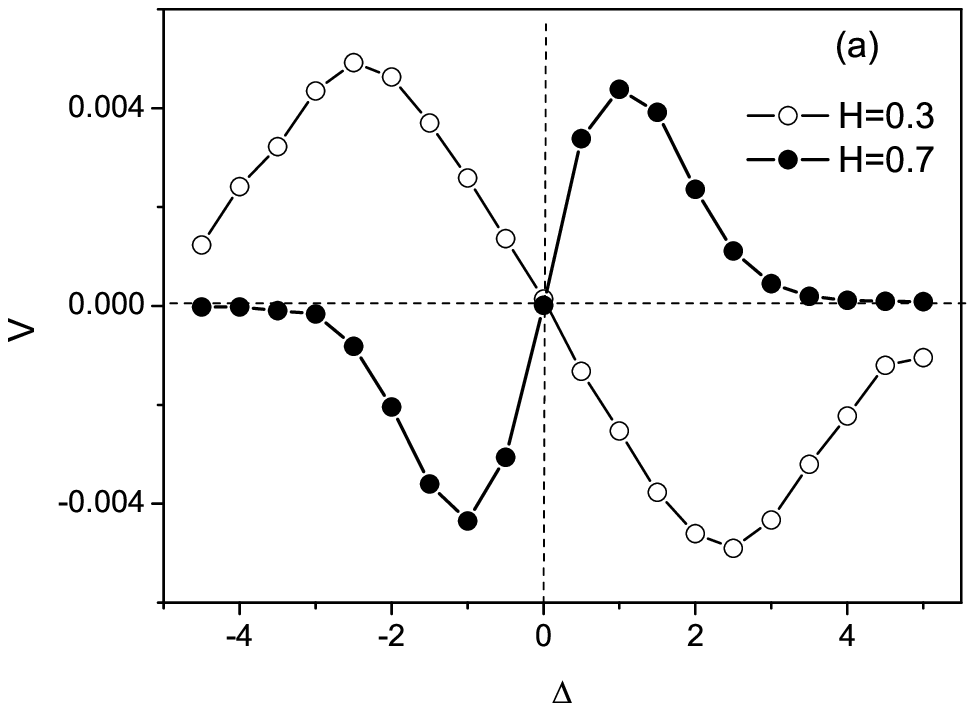}
  \includegraphics[width=8cm,height=8cm]{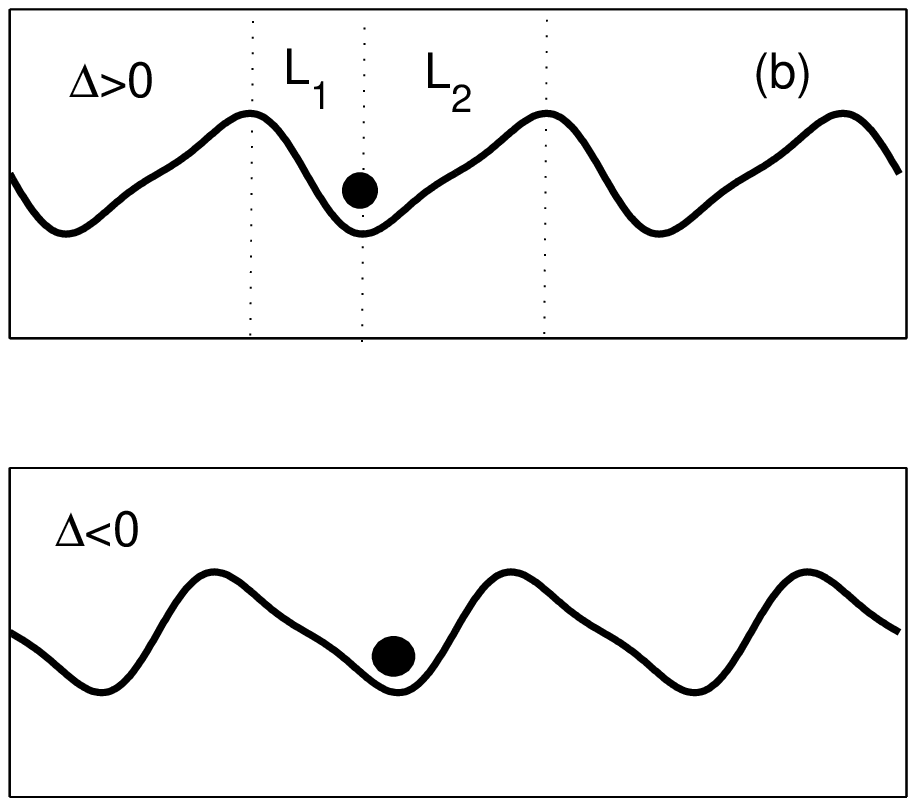}
  \caption{(a)Average velocity $v$ as a function of the asymmetric parameter $\Delta$ of the potential for both persistent and anti-persistent cases at $D=0.3$ and $U_{0}=1$.
  (b)Asymmetrically periodic potential for $\Delta >0$ and $\Delta<0$, $L_{1}$ is the length from the minima of the potential to the maxima
  from the left side and $L_{2}$ is the length from the right side.}\label{1}
\end{center}
\end{figure}
\indent Next, we will study the directed transport mechanism for our
ratchet. Usually, the ratchet mechanism demands three key
ingredients \cite{a1}: (a) nonlinear periodic potential,
(b)asymmetry of the potential or external driving forces, and
(c)fluctuating. Figure 4 (a) shows the average velocity as a
function of the asymmetry of the potential. For the persistent case
($H=0.7$), the velocity is positive for $\Delta>0$, zero at
$\Delta=0$, and negative for $\Delta<0$. However, for
anti-persistent case one can obtain the opposite velocity, negative
for $\Delta>0$ and positive for $\Delta<0$ . Moreover, for both
cases, there exists an optimal value of $\Delta$ at which the
velocity takes its extremal value. When $\Delta\rightarrow 0$, the
system is absolutely symmetric and directed transport disappears.
When $\Delta\rightarrow \infty$, the asymmetric potential described
in Eq. (2) reduces to symmetric one ($U(x)=-\frac{U_{0}}{4}\Delta
\sin(2x)$)with higher barriers,
resulting in zero velocity. \\
\indent Now we will give the physical interpretation of the directed
transport for the case of $\Delta=1$ (see the upper of Fig. 4 (b)).
We define three time periods that are very important for explanation
of the directed transport: driving period $T$, diffusion time
$T_{1}$ for crossing the steeper slope (the left side) from the
minima, and diffusion time $T_{2}$ for crossing the gentler
slope(the right side). Because of $L_{1}<L_{2}$, $T_{1}$ is always
less than $T_{2}$.  Firstly, the particles stay in the minima of the
potential (see Fig. 4 (b)) until they are catapulted out of the well
by a large amplitude fluctuation.  For the persistent case
($H=0.7$), the fractional Gaussian noise contains more low frequency
components (see the upper of the Fig. 3) and it can be divided into
two parts: white Gaussian noise and low frequency ac drive. Due to
the low frequency, the drive has a very long period and $T\gg
T_{2}>T_{1}$. All particles get enough time to cross both sides from
the minima of the potential before the drive reverses its direction.
However, the left side is steeper than the right one, more particles
climb the barrier from the right side, so the average velocity is
positive. For the anti-persistent case($H=0.3$), the high frequency
components dominate over the low frequency ones(see the bottom of
the Fig. 3). In this case, the drive has a short period and
$T_{1}<T<T_{2}$ (or $T<T_{1}<T_{2}$).  In a short driving period,
the particles have sufficient time to diffuse across the steeper
side of the well, resulting in negative average velocity. It should
be pointed out that the average velocity will tend to zero for the
case of $T\ll T_{1}<T_{2}$ (very small values of $H$) which is also
shown in Fig. 6.
\begin{figure}[htbp]
  \begin{center}\includegraphics[width=10cm,height=8cm]{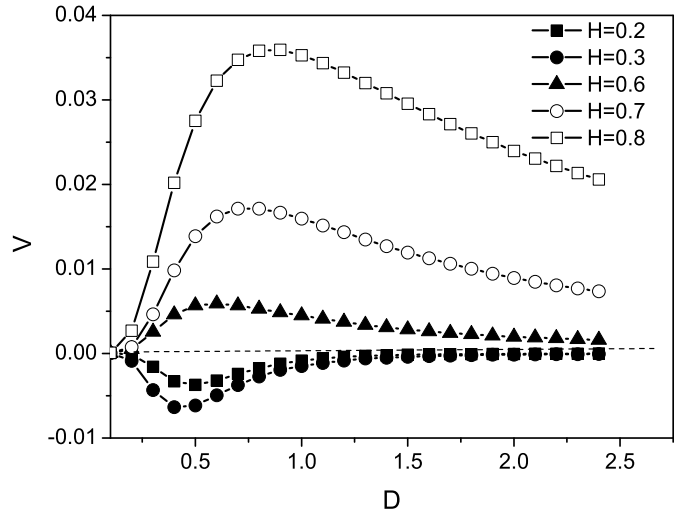}
  \caption{Average velocity $V$ as a function of noise intensity $D$ for different values of Hurst exponent $H=0.2$, $0.3$, $0.6$, $0.7$, and $0.8$ at
  $\Delta=1.0$ and $U_{0}=1$. }\label{1}
\end{center}
\end{figure}

\indent The noise intensity dependence of the average velocity is
shown in Fig. 5 for different values of Hurst exponent.  The curve
is observed to be bell shaped. When $D\rightarrow 0$, the particles
cannot pass across the barrier and there is no directed current.
When $D\rightarrow \infty$ so that the noise is very large, the
effect of the potential disappears and the average velocity tends to
zero, also. Therefore, one can see that the curves demonstrate
nonmonotonic behavior.

\begin{figure}[htbp]
  \begin{center}\includegraphics[width=10cm,height=8cm]{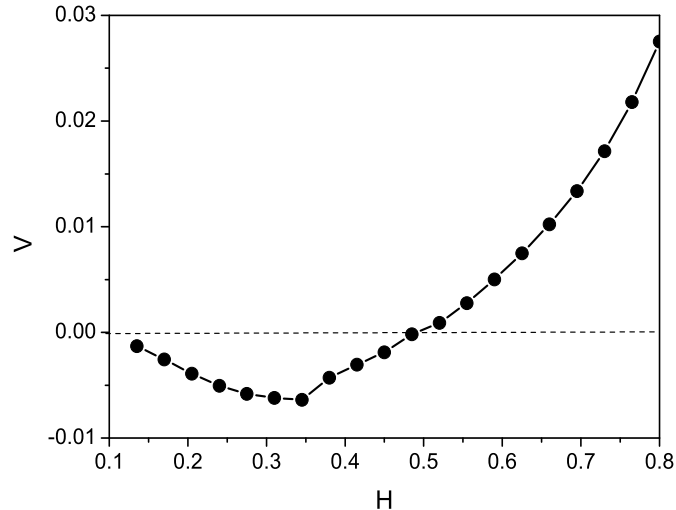}
  \caption{Average velocity $V$ as a function of the Hurst exponent $H$ at
  $\Delta=1.0$, $D=0.5$, and $U_{0}=1$.}\label{1}
\end{center}
\end{figure}

\indent Figure 6 displays the Hurst exponent dependence of the
average velocity at $\Delta=1.0$. It is found that the average
velocity is positive for $H>\frac{1}{2}$, zero at $H=\frac{1}{2}$,
and negative for $H<\frac{1}{2}$. For the persistent case
($H>\frac{1}{2}$), the average velocity increases monotonically with
the Hurst exponent. However, for the anti-persistent case
($H<\frac{1}{2}$), there exists a value of Hurst exponent at which
the average velocity takes its extremal value. When $H\rightarrow
0$, the noise term in Eq. (1) will disappear, the system is
deterministic and the directed transport also disappears. When
$H\rightarrow \frac{1}{2}$, the fractional Gaussian noise reduces to
the white Gaussian noise, the system undergoes thermal equilibrium
and the average velocity tends to zero.

\section{Concluding Remarks}
\indent In this paper, we studied the directed transport of
overdamped Brownian particles driven by fractional Gaussian noises.
From numerical simulations, we can find that fractional Gaussian
noise can break the detailed balance and induce directed transport.
Similar to the classic ratchets \cite{a1}, there exists a value of
noise intensity at which the average velocity takes its extremal
value. The average velocity as a function of the asymmetry of the
potential is monotonic. From the numerical analysis of the spectral
density of fractional Gaussian noises, it is found that the low
frequency component in spectral density is larger than the high
frequency component for the persistent case ($H>\frac{1}{2}$),
whereas the high frequency component is dominated for the
anti-persistent case ($H<\frac{1}{2}$). Due to the difference of the
spectral density between the persistent and anti-persistent cases,
the average velocity has the opposite sign for the two cases.
Remarkably, the average velocity increases monotonically with the
Hurst exponent for the persistent case. However, for anti-persistent
case, there exists an optimal value of the Hurst exponent at which
the velocity takes
its extremal value.\\
  \indent Directed transport in static ratchet potentials can also
  be induced by the other types of noise, such as  $\alpha$-stable
  noise (L\'{e}vy ratchet)\cite{a11,a12,a13,a14}, white shot noise (shot-noise ratchet)\cite{c1},
  and two correlated noises (correlated ratchet)\cite{a7}. In the L\'{e}vy
  ratchet \cite{a11,a12,a13,a14}, the thermodynamical equilibrium is broken by the
  the heavy-tailed distribution of the $\alpha$-stable
  noise.  For shot-noise ratchet \cite{c1}, the temporal asymmetry of white
  shot noise can induce an effective, inhomogeneous diffusion, so
  the net current occurs.  In the correlated ratchet \cite{a7},
  fluctuation-induced transport is driven by both additive Gaussian
  white noise and additive colored noise.  The additive colored noise
  can be treated as the multiplicative noise by introducing a new auxiliary variable,  therefore an effective, inhomogeneous
  diffusion appears. However, in our fractional Gaussian
  noise-induced ratchet, the directed transport is induced by the
  asymmetry of noise spectral density.  When $H>\frac{1}{2}$, the
  fractional Gaussian noise contains more low frequency components,
  whereas the high frequency component is dominated for
  $H<\frac{1}{2}$.

  \indent This work was supported in part by National
Natural Science Foundation of China (Grant Nos. 30600122, 11004982
and 10947166 )and GuangDong Provincial Natural Science Foundation
(Grant No. 06025073 and 01005249). Y. F. He also acknowledges the
Research Foundation of Education Bureau of Hebei Province, China
(Grant No. 2009108)


\begin{thebibliography}{}
\bibitem{a1}P. H\"{a}nggi and F. Marchesoni, Rev. Mod. Phys. 81, 387
(2009).
\bibitem{a2}F. Julicher A. Adjari, and J. Prost, Rev. Mod. Phys. 69, 1269 (1997).
\bibitem{a3}J. Rousselet, L. Salome, A. Adjari, and J. Prost, Nature 370, 446 (1994);
L. P. Faucheux, L. S. Bourdieu, P. D. Kaplan, and A. J. Libchaber,
Phys. Rev. Lett. 74, 1504 (1995).
\bibitem{a4}F. Marchesoni, Phys. Rev. Lett. 77, 2364(1996); I. Derenyi, C. Lee, and A. L.
Barabasi, Phys. Rev. Lett. 80,1473 (1998); C. S. Lee et al., Nature
400, 337 (1999).
\bibitem{a5} M. O. Magnasco, Phys. Rev. Lett. 71, 1477 (1993);
R. Bartussek, P. H\"{a}nggi, and J. G. Kissner, Europhys. Lett. 28,
459 (1994).
\bibitem{a6}P. Reimann, Phys. Rep. 290, 149(1997);
J. D. Bao and Y. Z. Zhuo, Phys. Lett. A 239, 228 (1998); B. Q. Ai,
L. Q. Wang, and L. G. Liu, Chaos, Solitons Fractals 34, 1265 (2007);
P. Reimann, R. Bartussek, R. Haussler, and P. H\"{a}nggi, Phys.
Lett. A 215, 26 (1996).
\bibitem{a7}C. R. Doering, W. Horsthemke, and J. Riordan, Phys. Rev. Lett. 72, 2984
(1994); R. Bartussek, P. Reimann, and P. H\"{a}nggi, Phys. Rev.
Lett. 76, 1166(1996).
\bibitem{a8}B. Q. Ai and L. G. Liu, Phys. Rev. E 74, 051114 (2006);
B. Q. Ai, Phys. Rev. E 80, 011113 (2009); F. Marchesoni, S.
Savel'ev, Phys. Rev. E 80, 011120 (2009); B. Q. Ai, J. Chem. Phys.
131, 054111 (2009).
\bibitem{a9}R. Metzler and J. Klafter, Phys. Rep. 339, 1 (2000);
 I. Goychuk and P. H\"{a}nggi, Phys. Rev. Lett. 99, 200601
(2007).
\bibitem{a10}B. Dybiec, E. Gudowska-Nowak, and I. M. Sokolov, Phys.
Rev. E 78, 011117 (2008); B. Dybiec, Phys. Rev. E 78, 061120 (2008).
\bibitem{a11}D. Del-Castillo-Negrete, V. YU. Gonchar, and A. V.
Chechkin, Phys. Lett. A 387, 6693 (2008).
\bibitem{a12}I. Goychuk, E. Heinsalu, M. Patriarca, G. Schmid,and P.
H\"{a}nggi, Phys. Rev. E 73, 020101(R) (2006); E. Heinsalu, M.
Patriarca, I. Goychuk, G. Schmid, and P. H\"{a}nggi, Phys. Rev. E
73, 046133 (2006).
\bibitem{a13}J. Rosa and M. W. Beims, Physica A 386, 54 (2007).
\bibitem{a14}B. Q. Ai and Y. F. He, J. Stat. Mech.: Theory Exp., P04010 (2010).
\bibitem{a15}B. B. Mandelbrot and J. W. Van Ness, SIAM Rev. 10, 422
(1968).
\bibitem{a16}D. Panja, J. Stat. Mech.: Theory Exp., L02001 (2010).
\bibitem{a17}G. Guigas and M. Weiss, Biophys. J. 94, 90 (2008);
J. Szymanski and M.Weiss, Phys. Rev. Lett. 103, 038102 (2009).
\bibitem{a18}L. Lizana and T. Ambjornsson, Phys. Rev. Lett. 100,
200601 (2008).
\bibitem{a19}A. Zoia, A. Rosso, and S. N. Majumdar, Phys. Rev. Lett. 102,
120602 (2009).

\bibitem{a20}K. Burnecki and A. Weron, Phys. Rev. E 82, 021130
(2010); A. Weron, K. Burnecki, Sz. Mercik, and K. Weron, Phys. Rev.
E 71, 016113 (2005).
\bibitem{a21}I. Calvo and R. Sanchez, J. Phys. A: Math. Theor. 41,
282002 (2008); M. Bologna, F. Vanni, A. Krokhin, and P. Grigolini,
Phys. Rev. E 82, 020102(R) (2010); L. Zunino, D. G. Perez, M. T.
Martin, M. Garavaglia, A. Plastino, and O. A. Rosso, Phys. Lett. A
372, 4768 (2008).
\bibitem{a22} T. A. {\O}igard, A. Hanssen, and L. L. Scharf, Phys. Rev. E
74, 031114 (2006); W. Deng and E. Barkai, Phys. Rev. E 79, 011112
(2009); J. H. Jeon and R. Metzler, Phys. Rev. E 81, 021103 (2010).
\bibitem{a23}R. Garcia-Garcia, A. Rosso, and G. Schehr, Phys. Rev. E
81, 010102(R) (2010); N. Kumar, U. Harbola, and K. Lindenberg, Phys.
Rev. E 82, 021101 (2010).
\bibitem{a24}M. Magdziarz, A. Weron, K. Burnecki, and J. Klafter,
Phys. Rev. Lett 103, 180602 (2009); I. Eliazar and J. Klafter, Phys.
Rev. E 79, 021115 (2009).
\bibitem{a25}O. Y. Sliusarenko, V. Y. Gonchar, A. V. Chechkin, I. M.
Sokolov, and R. Metzler, Phys. Rev. E 81, 041119, (2010); O. Y.
Sliusarenko, V. Y. Gonchar, and A. V. Checkin, Urk. J. Phys. 55, 579
(2010).
\bibitem{a26}S. Chaudhury and B. J. Cherayila, J. Chem. Phys. 125,
114106 (2006).
\bibitem{b1} S. C. kou and X. S. Xie, Phys. Rev. Lett. 93, 180603
(2004).
\bibitem{b2}A. V. Checkin and V. Y. Gonchar, Chaos, Solition and
Fractals 12, 391(2001) ; B.S. Lowen, Meth. Comput. Applied Probab.
1:4, 445 (1999).
\bibitem{b3}T. Dieker, \emph{Simulation of Fractional Brownian Motion}, Masters
Thesis, Department of Mathematical Sciences, University of Twente,
The Netherlands, 2004.
\bibitem{c1}J. Luczka, R. Bartussek, and P. H\"{a}nggi, Europhys.
Lett. 31, 431 (1995).

\end{thebibliography}
\end{document}